\documentclass[%
 reprint,
 twocolumn,
 amsmath,amssymb,
 aps, 
prl,
]{revtex4-2}

\usepackage{graphicx}
\usepackage{dcolumn}
\usepackage{bm}
\usepackage{lipsum}

\usepackage{orcidlink}
\usepackage[dvipsnames]{xcolor}
\usepackage{labelfig}
\usepackage{mathrsfs}
\usepackage{hyperref}
\usepackage{braket}

\newcommand{\qubit}[1]{$\lvert#1\rangle$}

\begin{document}

\preprint{APS/123-QED}

\title{\textbf{Quantum Enhancement of Particle-Size Segregation} 
}%

\author{Tomás Trewhela}
 \email{Contact author: tomas.trewhela@uai.cl}
\affiliation{Facultad de Ingeniería y Ciencias, Universidad Adolfo Ibáñez, Viña del Mar, Chile.}

\date{\today}

\begin{abstract}
Segregated states based on particle size emerge in granular materials from the competition between segregation and diffusive remixing. Here, we show that quantum coherence can enhance segregation beyond this classical limit. We introduce an open quantum cellular automaton for bidisperse mixtures that combines coherent transport and dissipative segregation. The automaton reproduces experimental and continuum-theory segregation dynamics, with segregation degrees collapsing onto a theoretical Péclet-dependent relationship. However, weakly decohering systems exhibit a coherence-driven transport regime that produces more strongly segregated steady states than classical predictions. Across a broad parameter range, the steady-state degree of segregation collapses onto two dimensionless numbers governing the competition between segregation, diffusion, and decoherence. These results identify quantum coherence as a mechanism for enhancing particle-size segregation and establish a framework for studying transport phenomena in open many-body systems.
\end{abstract}

\maketitle


A collection of small and large grains subjected to shear under gravity tends to segregate according to particle size \cite{Ottino00,Gray18,Umbanhowar19}. Gravity-driven segregation percolates small particles \cite{Middleton70} to the bottom and drives large particles toward the free surface \cite{Rosato87,Savage88}, while diffusion promotes remixing and opposes the formation of ordered states \cite{Gray18,Thornton26}. The competition between these mechanisms governs the evolution of granular mixtures and ultimately explains the many fascinating steady-state grain patterns observed in a wide range of natural and industrial flows \cite{Makse97a,Johnson12,Baker16b,Pearse26}.

Several approaches have been developed to model particle-size segregation, among which continuum mixture-theory formulations have become the dominant framework over the past two decades \cite{Gray18,Umbanhowar19,Thornton26}. These models describe the evolution of particle concentrations through segregation-diffusion equations that accurately predict concentration fields in space and time. Complementary descriptions have also been developed using cellular automata, in which local particle rearrangements collectively reproduce macroscopic segregation dynamics when averaged \cite{Fitt92,Yanagita99,Marks11,Castro22,Dissanayake25}. By representing transport through local update rules rather than continuum fluxes, cellular automata provide a natural framework for exploring alternative mechanisms capable of modifying the classical segregation-diffusion balance \cite{Baxter90}. 

Classical analogues of quantum phenomena are frequently encountered in condensed matter and statistical physics, providing valuable insights into quantum behavior through accessible macroscopic systems \cite{Longhi09,Bush15}. 
Much less explored is whether quantum concepts themselves can reveal new transport mechanisms in systems traditionally regarded as entirely classical. Recent advances in open quantum dynamics have shown that the interplay between coherence and dissipation can produce transport behavior beyond classical predictions \cite{Breuer02,Plenio08,Caruso09}. In the context of particle-size segregation, a quantum framework provides a natural setting for exploring how coherent transport modifies the classical segregation-diffusion balance while retaining dissipative descriptions of local particle rearrangements.

In this letter, we first introduce the framework consisting of an open quantum cellular automaton for bidisperse particle segregation. The automaton represents a one-dimensional granular column as a lattice of two-state qubits corresponding to small and large grains \cite{Nielsen10}. Coherent mixing and gravitational gates combine with dissipative Lindblad jump operators to produce segregation, diffusion, coherence, and decoherence within a unified framework \cite{Manzano20}. We show that the quantum automaton, in a single run, recovers the classical segregation-diffusion dynamics. Moreover, it yields similar steady-state solutions to those of continuum theory numerical solutions, with the automaton degrees of segregation collapsing onto an analogous Péclet number. Beyond this classical limit, however, the automaton reveals an intriguing coherence-driven transport regime that produces higher degrees of segregation than classical predictions. Furthermore, the dynamics collapse onto a second dimensionless quantity governed by the competition between coherent segregation and decoherence.

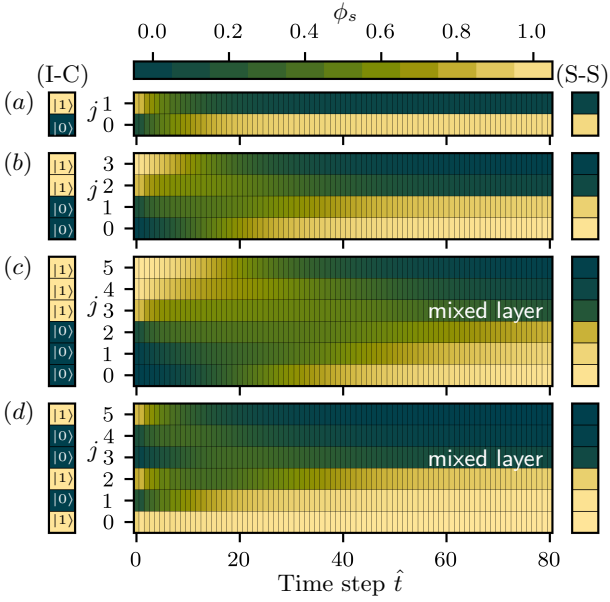
\begin{figure}[!h]
\begin{center}
\SetLabels
\L (0.05*0.875) (I-C)\\
\L (0.0*0.83) ($a$)\\
\L (0.13*0.815) $j$\\
\L (0.075*0.83) \tiny\qubit{1}\\
\L (0.075*0.79) \textcolor{white}{\tiny\qubit{0}}\\
\L (0.0*0.725) ($b$)\\
\L (0.13*0.68) $j$\\
\L (0.075*0.725) \tiny\qubit{1}\\
\L (0.075*0.687) \tiny\qubit{1}\\
\L (0.075*0.65) \textcolor{white}{\tiny\qubit{0}}\\
\L (0.075*0.615) \textcolor{white}{\tiny\qubit{0}}\\
\L (0.0*0.545) ($c$)\\
\L (0.65*0.47) \textcolor{white}{\sf mixed layer}\\
\L (0.65*0.215) \textcolor{white}{\sf mixed layer}\\
\L (0.13*0.48) $j$\\
\L (0.075*0.545) \tiny\qubit{1}\\
\L (0.075*0.51) \tiny\qubit{1}\\
\L (0.075*0.475) \tiny\qubit{1}\\
\L (0.075*0.44) \textcolor{white}{\tiny\qubit{0}}\\
\L (0.075*0.404) \textcolor{white}{\tiny\qubit{0}}\\
\L (0.075*0.367) \textcolor{white}{\tiny\qubit{0}}\\
\L (0.0*0.295) ($d$)\\
\L (0.13*0.23) $j$\\
\L (0.075*0.295) \tiny\qubit{1}\\
\L (0.075*0.26) \textcolor{white}{\tiny\qubit{0}}\\
\L (0.075*0.224) \textcolor{white}{\tiny\qubit{0}}\\
\L (0.075*0.187) \tiny\qubit{1}\\
\L (0.075*0.152) \textcolor{white}{\tiny\qubit{0}}\\
\L (0.075*0.115) \tiny\qubit{1}\\
\L (0.505*0.985) $\phi_{s}$\\
\L (0.42*0.00) Time step $\hat{t}$\\
\L (0.845*0.875) (S-S)\\
\endSetLabels
\strut\AffixLabels{\includegraphics[width=\columnwidth]{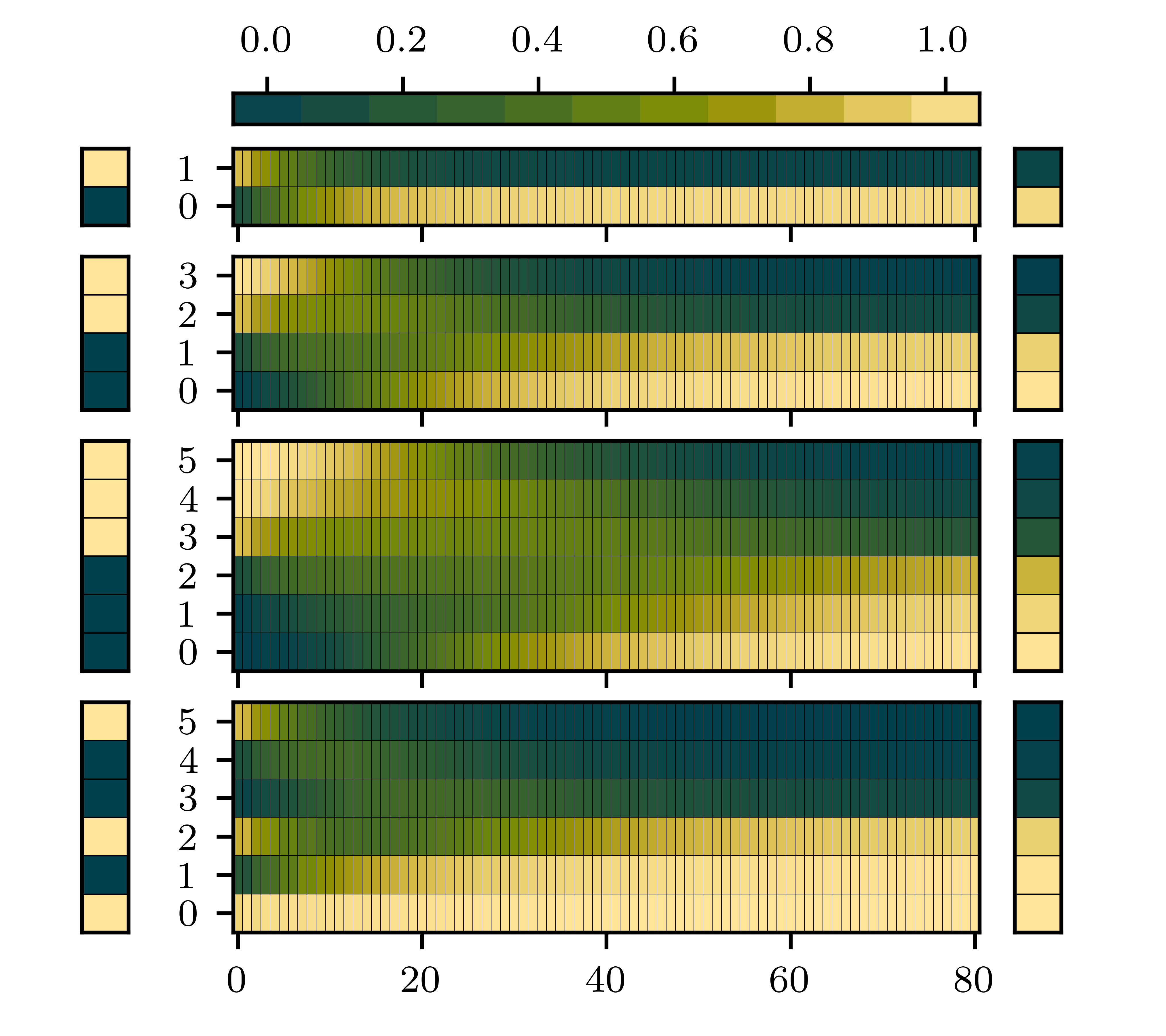}}
\end{center}
\vspace{-0.5cm}
\caption{($a$--$c$) Quantum cellular automaton simulations of the small-particle concentration $\phi_s(j,\hat{t})$ for an initially unstable normally graded configuration (I-C) and lattice sizes $N={2,4,6}$, respectively. ($d$) Simulation initialized from a randomly mixed configuration. In all cases, the automaton evolves toward an inversely graded steady state (S-S), characterized by large particles near the free surface \cite{Rosato87,Savage88} and small particles at depth\cite{Middleton70}, consistent with gravity-driven particle-size segregation\cite{Umbanhowar19,Thornton26} The steady-state profiles exhibit finite mixed layers resulting from the competition between segregation and remixing \cite{Gray06,Ulloa26}.}
\label{fig:QCA}
\end{figure}

\emph{Quantum cellular automaton.} We consider a one-dimensional granular column composed of small and large particles of diameters $d_s$ and $d_l>d_s$, respectively, represented by a lattice of $N$ sites (Fig.~\ref{fig:QCA}I-C). Each site is occupied by either a small or a large particle, represented by the local basis states \qubit{s}$\equiv$\qubit{1} and \qubit{l}$\equiv$\qubit{0}, respectively. The state of the system is described by a density matrix $\rho$ acting on the Hilbert space
$\mathcal{H}=(\mathbb{C}^2)^{\otimes N}$. Starting from a specific initial arrangement, the system evolves through local particle rearrangements driven by mixing and size segregation. Since these rearrangements occur only between neighboring sites, the elementary dynamics are defined on the two-site Hilbert space $\mathbb C^2\otimes\mathbb C^2$, spanned by the basis states ${\ket{ab}}_{a,b\in{0,1}}$.

Mixing is represented by coherent exchanges between neighboring particles of different species. These exchanges are generated by the Hamiltonian $H_{\mathcal D}=\ket{sl}\bra{ls}+\ket{ls}\bra{sl}$ with the corresponding unitary evolution $U_{\mathcal D}=e^{-i\delta H_{\mathcal D}}$, where $\delta$ controls the strength of coherent mixing. The operator acts on all possible pairs, but it only has a net effect on the heterogeneous pairs $\ket{sl}$ and $\ket{ls}$, leaving homogeneous pairs unchanged. To introduce a preferred vertical direction, we include the gravitational Hamiltonian $H_g=g\sum_j z_j n^l_j$, where $n^{l}_j$ denotes the large-particle occupation at height $z_j$, and $g=1$ induces the direction and bias for the segregation direction. The corresponding unitary evolution $U_g=e^{-iH_g\Delta\hat t}$ provides the gravitational bias required for segregation \cite{Gray18,Trewhela24b}. 

Particle-size segregation, in contrast, is an irreversible process arising from kinetic sieving and squeeze expulsion \cite{Middleton70,Savage88,Gray18,Umbanhowar19}. We represent these mechanisms through the Lindblad jump operator $\mathcal L_{sl}=\sqrt{\Gamma}\,\ket{sl}\bra{ls}$, where $\Gamma$ is the segregation-controlling parameter. The operator transforms the unstable configuration $\ket{ls}$ into the stable configuration $\ket{sl}$, thereby driving local segregation. To account for decoherence, we introduce local dephasing operators $\mathcal L_{\phi}^{(j)}=\sqrt{\gamma}\,\sigma_z^{(j)}$,
where $\gamma$ is the dephasing rate and $\sigma_z^{(j)}$ acts on lattice site $j$. These operators suppress coherent correlations between particle configurations without altering particle concentrations, thereby driving the automaton toward the classical segregation-diffusion limit in which transport is governed solely by segregation and diffusive remixing.

With all these, the competition between coherent mixing, dissipative segregation, and dephasing is described by the Lindblad master equation \cite{Nielsen10,Manzano20}
\begin{equation}
    \frac{\text{d}\rho}{\text{d}\hat{t}}=-\frac{i}{\hbar}[H,\rho]+\sum_j\left(L_j\rho L_j^\dagger-\frac{1}{2}\{L_j^\dagger L_j,\rho\}\right),
\label{eq:lindblad}
\end{equation}
where $H=H_\mathcal{D}+H_{g}$ generates coherent transport, and $L_j$ denotes the set of segregation and dephasing operators. Equation~\ref{eq:lindblad} therefore recasts segregation-diffusion dynamics as an open quantum transport process in which coherent mixing competes with segregation and decoherence. In the strong-dephasing limit, quantum coherences are rapidly suppressed, and the automaton approaches the classical segregation-diffusion regime.

To characterize segregation, we define the local occupation operator $\hat n_j^\nu=\ket{\nu}_j\bra{\nu}$ for particle species $\nu\in\{s,l\}$. The local concentration is then given by $\phi_\nu(j,t)=\mathrm{Tr}\left[\rho(t)\hat n_j^\nu
\right]$. For a bidisperse system, conservation of occupancy yields $\phi_s(j,t)+\phi_l(j,t)=1$, allowing direct comparison with continuum segregation models \cite{Gray18}.

Figure~\ref{fig:QCA} presents quantum cellular automaton simulations for lattice sizes $N={2,4,6}$ and two initial conditions (Fig.~\ref{fig:QCA}I-C): a normally graded configuration (Fig.~\ref{fig:QCA}$c$) and a randomly distributed (mixed) configuration (Fig.~\ref{fig:QCA}$d$). In both cases, the automaton evolves toward particle-size-segregated states consistent with the classical segregation-diffusion framework. Starting from an initial density matrix $\rho(0)$, coherent mixing generated by $U_{\mathcal D}$ induces reversible exchanges between neighboring particles, while the Lindblad segregation operators drive irreversible local rearrangements. The competition between these processes produces concentration profiles characteristic of granular segregation. The presence of a persistent mixed layer (in Fig.~\ref{fig:QCA}S-S) is a hallmark of segregation-diffusion balances \cite{Gray06,Wiederseiner11,Gray18,Pearse26,Ulloa26} and demonstrates that the automaton does not evolve toward complete segregation despite the irreversible nature of the segregation jumps.

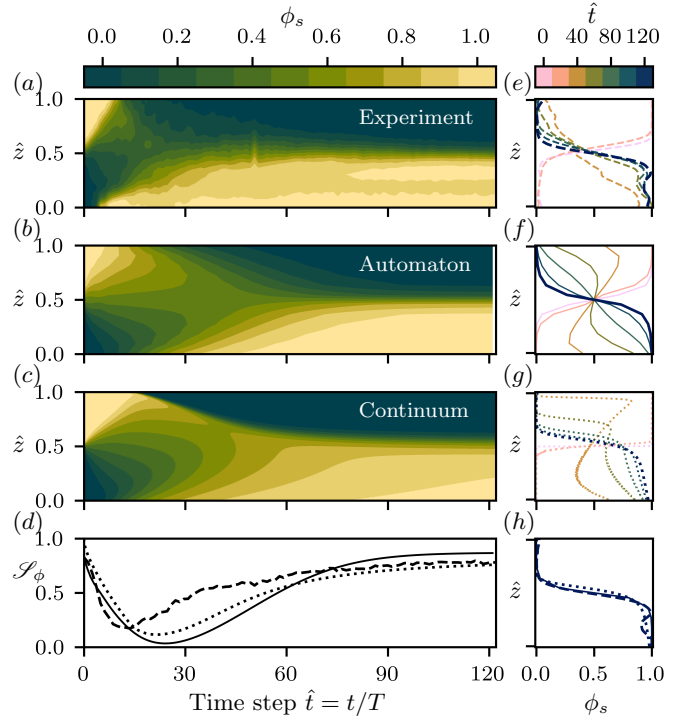
\begin{figure}[!h]
\begin{center}
\SetLabels
\L (0.0*0.89) ($a$)\\
\L (0.0*0.79) $\hat z$\\
\L (0.0*0.68) ($b$)\\
\L (0.0*0.58) $\hat z$\\
\L (0.0*0.47) ($c$)\\
\L (0.0*0.37) $\hat z$\\
\L (0.0*0.26) ($d$)\\
\L (0.0*0.19) $\mathscr{S}_{\phi}$\\
\L (0.75*0.89) ($e$)\\
\L (0.76*0.79) $\hat z$\\
\L (0.75*0.68) ($f$)\\
\L (0.76*0.58) $\hat z$\\
\L (0.75*0.47) ($g$)\\
\L (0.76*0.37) $\hat z$\\
\L (0.75*0.26) ($h$)\\
\L (0.76*0.16) $\hat z$\\
\L (0.41*0.985) $\phi_{s}$\\
\L (0.53*0.84) \textcolor{white}{\footnotesize Experiment}\\
\L (0.53*0.63) \textcolor{white}{\footnotesize Automaton}\\
\L (0.53*0.42) \textcolor{white}{\footnotesize Continuum}\\
\L (0.88*0.985) $\hat{t}$\\
\L (0.27*0.00) Time step $\hat{t}=t/T$\\
\L (0.875*0.00) $\phi_{s}$\\
\endSetLabels
\strut\AffixLabels{\includegraphics[width=\columnwidth]{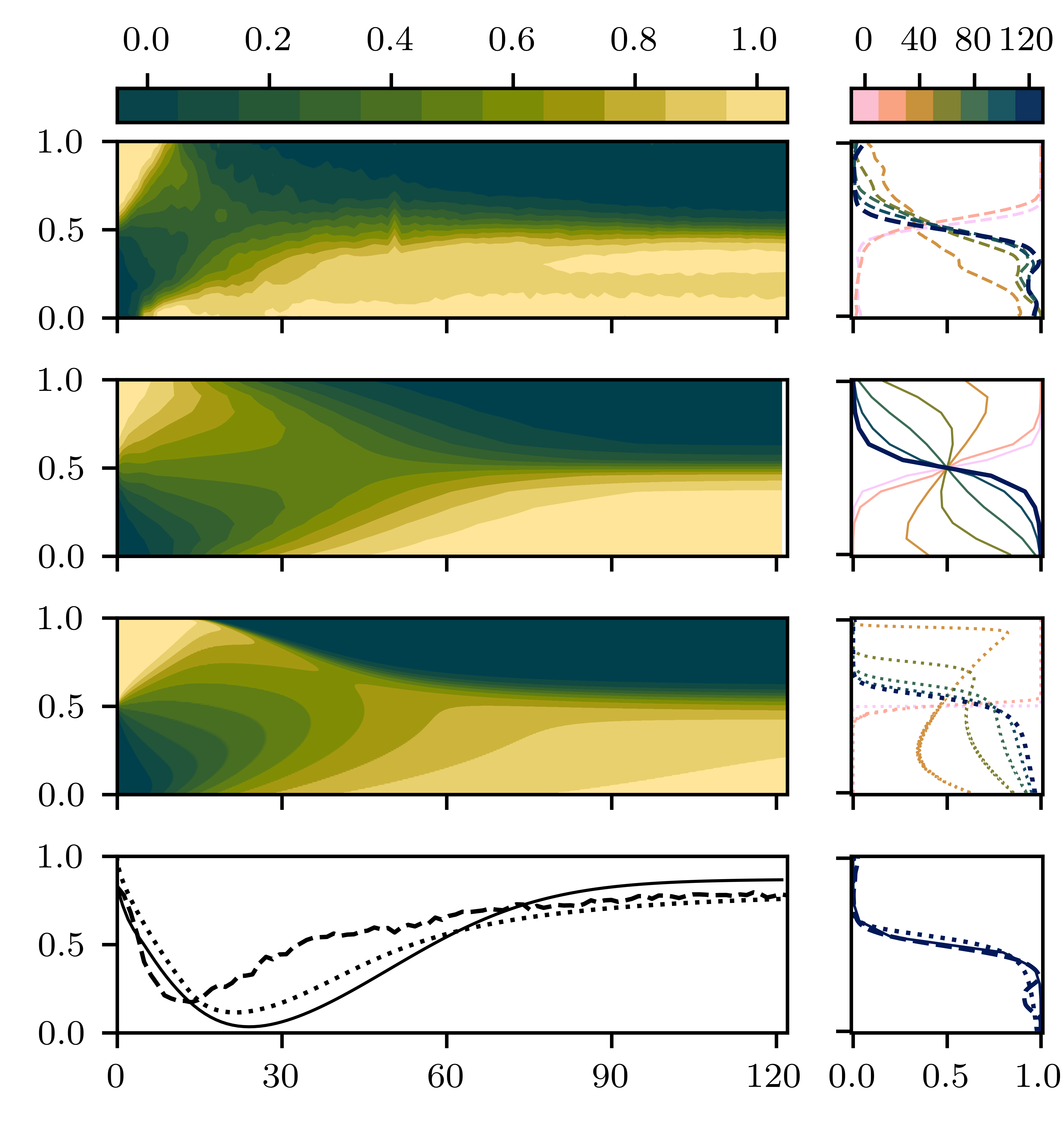}}
\end{center}
\vspace{-0.5cm}
\caption{($a$) Evolution of the small-particle concentration $\phi_s$ measured in the 50:50 oscillatory shear-cell experiment of Ref. \cite{vanderVaart15}. ($b$) Quantum cellular automaton simulation with $N=12$ and $\hat{t}=t/T=122$, corresponding to the 122 shear cycles of period $T$ in the experiment. ($c$) Numerical solution of the segregation-diffusion equation obtained using a method-of-lines (MOL) scheme \cite{Trewhela24a} with the experimental scaling law of Ref. \cite{Trewhela21a}. ($d$) Degree of segregation $\mathscr{S}_{\phi}$ \cite{Danckwerts52,Trewhela24b} as a function of nondimensional time $\hat{t}$ for the experiment (dashed), quantum cellular automaton (solid), and MOL solution (dotted). ($e$--$g$) Small-particle concentration profiles $\phi_s$ at selected times $\hat{t}$ for the experiment, automaton, and continuum model. ($h$) Comparison of the steady-state concentration profiles, demonstrating agreement between the experiment, the quantum cellular automaton, the continuum segregation-diffusion model, and the corresponding analytical steady-state solution \cite{Gray06,Trewhela21a,Trewhela24a}.}
\label{fig:comparison}
\end{figure}

\emph{Experimental and theoretical comparison.} Building on the qualitative segregation behavior shown in Fig.~\ref{fig:QCA}, we now benchmark the quantum cellular automaton against the oscillatory shear-cell experiments of Ref. \cite{vanderVaart15} and numerical solutions of the continuum segregation-diffusion equation based on mixture theory \cite{Gray18}. Figure~\ref{fig:comparison}($a$) shows the experimentally measured evolution of the small-particle concentration $\phi_s$ over successive shear cycles. Since concentration profiles were recorded once per oscillation period $T=13$, we compare these data directly with the automaton using the nondimensional time $\hat t=t/T$, which corresponds to one automaton time step. The automaton simulations were performed using $\Gamma=0.4$, $\delta=0.25$, and a lattice size of $N=12$ (Fig.~\ref{fig:comparison}$b$). For comparison, we also solved the one-dimensional segregation-diffusion equation using the method-of-lines implementation of Ref. \cite{Trewhela24a} together with experimentally calibrated segregation laws \cite{Trewhela21a,Trewhela21b}. Despite its minimal construction and the absence of an explicit segregation asymmetry\cite{vanderVaart15,Trewhela21a,Trewhela24a}, the automaton reproduces the main features of both the experiments and the continuum model. The connection between the Lindblad dynamics of the automaton and the continuum segregation--diffusion equation is outlined in the Supplemental Material. In particular, the temporal evolution of the degree of segregation $\mathscr{S}_\phi=1-(\sigma_\phi/\sigma_0)^2$, where $\sigma_\phi$ is the standard deviation of the concentration profile and $\sigma_0=0.5$ is the standard deviation of the fully segregated profile, shows that the automaton converges toward the same segregated attractor observed experimentally and predicted by continuum theory. This agreement becomes especially evident at steady state, where the concentration profiles obtained from all three approaches are nearly indistinguishable and agree with the analytical steady-state solution $\phi_{s}=e^{Pe(1/2-\hat{z})}/(1+e^{Pe(1/2-\hat{z})})$ \cite{Gray06,Wiederseiner11,Trewhela21a,Trewhela24a}.

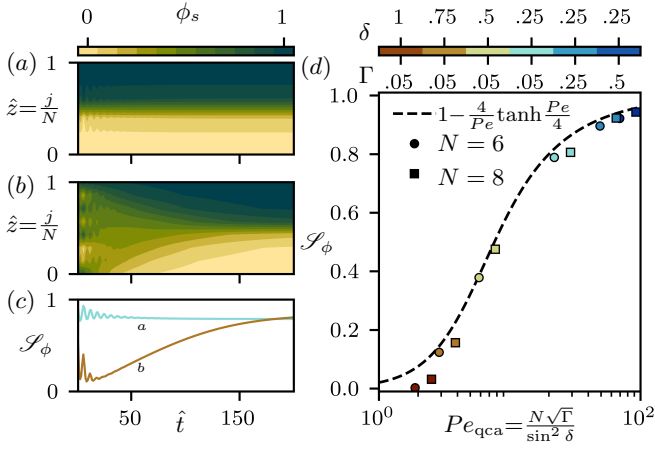
\begin{figure}[!h]
\begin{center}
\SetLabels
\L (0.0*0.85) ($a$)\\
\L (0.0*0.745) $\hat z\mathord{=}\frac{j}{N}$\\
\L (0.0*0.56) ($b$)\\
\L (0.0*0.47) $\hat z\mathord{=}\frac{j}{N}$\\
\L (0.0*0.29) ($c$)\\
\L (0.2*0.24) {\tiny$a$}\\
\L (0.2*0.14) {\tiny$b$}\\
\L (0.02*0.19) $\mathscr{S}_{\phi}$\\
\L (0.54*0.92) $\delta$\\
\L (0.54*0.82) $\Gamma$\\
\L (0.45*0.84) ($d$)\\
\L (0.45*0.43) $\mathscr{S}_{\phi}$\\
\L (0.66*0.73) {\footnotesize $1\mathord{-}\frac{4}{Pe}\text{tanh}\frac{Pe}{4}$}\\
\L (0.66*0.66) $N=6$\\
\L (0.66*0.58) $N=8$\\
\L (0.26*0.97) $\phi_s$\\
\L (0.26*0.00) $\hat{t}$\\
\L (0.67*0.00) $Pe_{\rm qca}\mathord{=}\frac{N\sqrt{\Gamma}}{\sin^{2}\delta}$\\
\endSetLabels
\strut\AffixLabels{\includegraphics[width=\columnwidth]{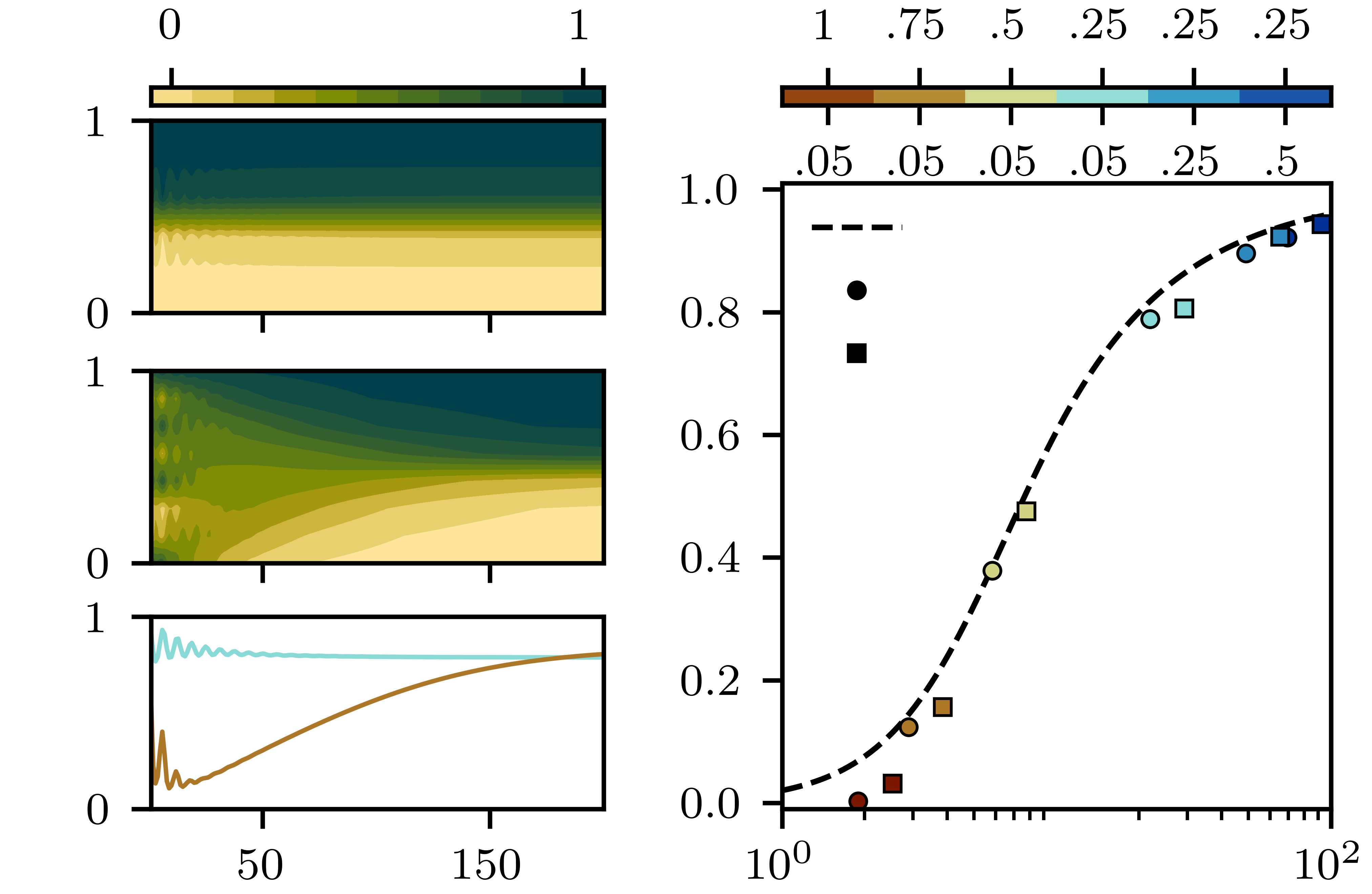}}
\end{center}
\vspace{-0.5cm}
\caption{($a$) Evolution of the small-particle concentration $\phi_s$ for an initially inversely graded configuration with $N=6$, $\delta=0.25$, $\gamma=0.01$, and $\Gamma=0.05$. ($b$) Evolution of $\phi_s$ for an initially mixed configuration with $N=8$, $\delta=0.25$, $\gamma=0.01$, and $\Gamma=0.05$. ($c$) Corresponding degrees of segregation $\mathscr{S}\phi$ for the simulations in ($a$) and ($b$), highlighting the initial coherence-driven transport regime prior to decoherence. ($d$) Steady-state degree of segregation $\mathscr{S}^{\infty}_\phi$ as a function of the segregation Péclet number $Pe_{\rm qca}=N\sqrt{\Gamma}/\sin^2\delta$ for simulations spanning $\delta={0.25,0.5,0.75,1.0}$, $\Gamma={0.05,0.25,0.5}$, and $N={6,8}$, with $\gamma=0.01$. The collapse demonstrates that the automaton recovers segregation-diffusion balance despite the slight presence of coherent transport.}
\label{fig:sdegree}
\end{figure}
A consequence of the agreement with the steady-state concentration profiles and segregation degrees shown in Fig.~\ref{fig:comparison}($d,h$) is that the automaton can be interpreted within the framework of segregation theory. We therefore examine whether the steady-state segregation degree acts as an attractor independent of the initial condition. Although preliminary evidence was already presented in Fig.~\ref{fig:QCA}, we now consider both inversely graded and randomly mixed initial states.

\emph{Classical degree of segregation scaling.} Figure~\ref{fig:sdegree}($a$) shows a simulation initialized from a fully inversely graded configuration consisting of three large particle sites above three small particle sites ($N=6$). Despite the initially sharp interface, the system evolves toward a steady state with a finite mixed layer, consistent with segregation-diffusion theory \cite{Gray06,Ferdowsi17,Ulloa26}. Differently, Fig.~\ref{fig:sdegree}($b$) shows a simulation initialized from a randomly mixed configuration. Although the transient evolution differs, the system again converges toward a diffuse segregation profile. The emergence of mixed layers contrasts with conventional cellular automata, which typically produce sharp segregation interfaces \cite{Marks11,Dissanayake25}.

For weak dephasing ($\gamma=0.01$), both simulations exhibit oscillations in the concentration field $\phi_s$ and the segregation degree $\mathscr{S}_\phi$ (Fig.~\ref{fig:sdegree}$c$). These oscillations arise from coherent transport and are progressively damped by decoherence. Nevertheless, the degree of segregation converges toward the same steady-state value regardless of the initial condition, demonstrating the existence of an attractor analogous to that observed in continuum segregation models \cite{Gray06,Trewhela24a,Trewhela24b}. The convergence toward a unique steady-state degree of segregation suggests the existence of a dimensionless parameter governing the balance between segregation and remixing. Motivated by the segregation parameter $\Gamma$, the coherent exchange probability $\sin^2\delta$, and the system size $N$, we define the quantum segregation Péclet number $Pe_{\rm qca}=N\sqrt{\Gamma}/\sin^2\delta$, which is analogous to the segregation Péclet number employed in continuum theories \cite{Fan14,Gray18,Umbanhowar19} (for its derivation, see the Supplemental Material). When the steady-state segregation degree is plotted against $Pe_{\rm qca}$, data obtained for different values of $\Gamma$, $\delta$, and $N$ collapse onto a single curve (Fig.~\ref{fig:sdegree}$d$).

\begin{figure}[!h]
\begin{center}
\SetLabels
\L (0.0*0.87) ($a$)\\
\L (0.13*0.98) Dephasing $\gamma$\\
\L (0.0*0.73) $\mathscr{S}_\phi$\\
\L (0.24*0.45) $\hat{t}$\\
\L (0.0*0.48) ($b$)\\
\L (0.0*0.32) $\mathscr{S}^{\infty}_\phi$\\
\L (0.45*0.85) ($c$)\\
\L (0.45*0.92) Diff. $\delta$\\
\L (0.45*0.78) Seg. $\Gamma$\\
\L (0.43*0.43) $\mathscr{S}^{\infty}_{\phi}$\\
\L (0.65*0.695) {\footnotesize $1\mathord{-}\frac{4}{Pe}\text{tanh}\frac{Pe}{4}$} \\
\L (0.62*0.63) $\gamma\mathord{=}0$\\
\L (0.62*0.55) $\gamma\mathord{=}1$\\
\L (0.13*0.00) Dephasing $\gamma$\\
\L (0.67*0.00) $Pe_{\rm qca}\mathord{=}\frac{N\sqrt{\Gamma}}{\sin^{2}\delta}$\\
\endSetLabels
\strut\AffixLabels{\includegraphics[width=\columnwidth]{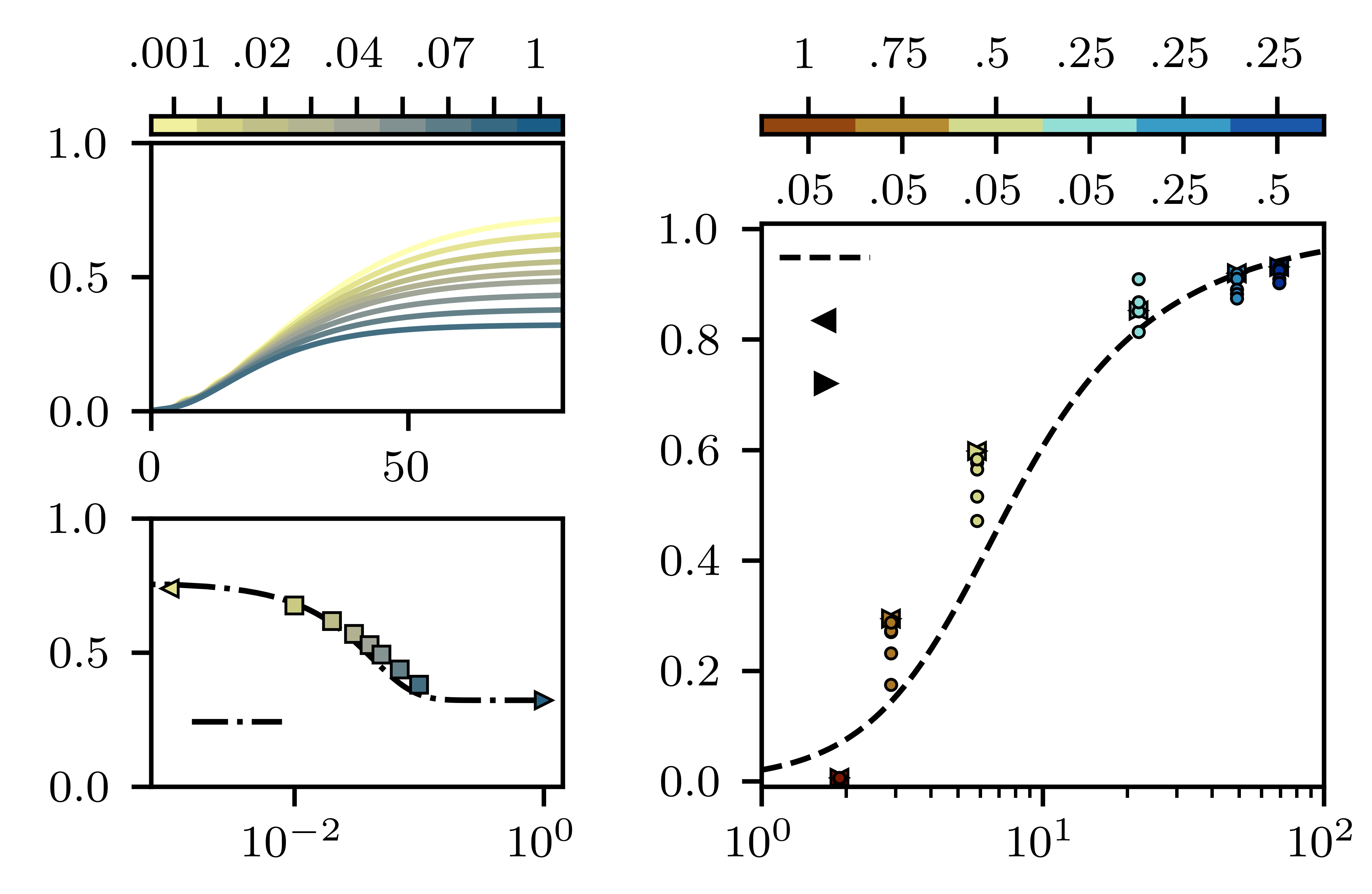}}
\end{center}
\vspace{-0.5cm}
\caption{($a$) Temporal evolution of the degree of segregation $\mathscr{S}_\phi$ for automaton simulations with different dephasing rates $\gamma$, showing the transition from coherence-enhanced segregation at weak dephasing to the classical segregation--diffusion regime at strong dephasing. ($b$) Steady-state segregation degree $\mathscr{S}_\phi^{\infty}$ as a function of the dephasing rate $\gamma$. Symbols denote simulation results and the dash-dotted line corresponds to the scaling relation given by Eq.~\eqref{eq:Scoh}. ($c$) Steady-state degree of segregation $\mathscr{S}_\phi^{\infty}$ as a function of the quantum cellular automaton Péclet number, including all simulations with varying $\Gamma$, $\delta$, and $\gamma$. The symbols collapse onto the theoretical segregation relationship $\mathscr{S}_\phi^{\infty}=1-\frac{4}{Pe}
\tanh\left(\frac{Pe}{4}\right)$ when the coherence-enhanced correction of Eq.~\eqref{eq:Scoh} is applied, demonstrating that the combined effects of segregation, diffusion, and decoherence can be described through the proposed scaling framework.}
\label{fig:coherence}
\end{figure}
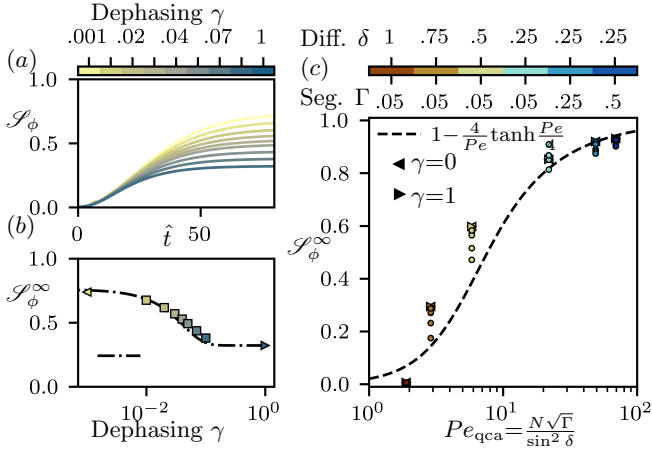

\emph{Coherence-enhanced segregation.} To isolate the role of coherence, we initialize the automaton in the superposed state $|\psi_0\rangle=\frac{1}{\sqrt2}\left(|\mathrm{inverse}\rangle+i|\mathrm{graded}\rangle\right)$, where $|\mathrm{inverse}\rangle$ and $|\mathrm{graded}\rangle$ denote the fully inverse-graded and graded configurations, respectively. Although this state produces a spatially uniform concentration profile, its density matrix contains long-range coherences between two macroscopically distinct segregation states. Consequently, any enhancement of segregation relative to the classical limit must originate from coherent transport rather than from classical segregation fluxes. 

Figure~\ref{fig:coherence}($a$) shows the evolution of the segregation degree for different dephasing rates $\gamma$. As $\gamma$ decreases, the trajectories progressively depart from the classical limit and converge toward increasingly segregated steady states. The strong-dephasing regime ($\gamma=1$) recovers the classical segregation-diffusion attractor described by $Pe_{\rm qca}$, whereas weakly dephased systems exhibit a distinct coherence-enhanced segregation regime.

To quantify this behavior, we introduce the dephasing number $\Pi_d=N\gamma/\sin^2\delta$, which balances dephasing and coherent transport timescales (for its derivation, see the Supplemental Material). Combining $\Pi_d$ with the quantum segregation Péclet number yields $\Pi_c=Pe_{\rm qca}/\Pi_{d}=\sqrt{\Gamma}/\gamma$, from which we derive a scaling for the coherence-enhanced degree of segregation 
\begin{equation}
    \mathscr{S}_\phi^{\infty}=\left[\frac{4\gamma}{\sqrt{\Gamma}}\,\text{tanh}\left(\frac{\sqrt{\Gamma}}{4\gamma}\right)\right]\Delta\mathscr{S}+\mathscr{S}^{\infty}_\phi|_{\gamma=1},
    \label{eq:Scoh}
\end{equation}
where $\Delta\mathscr{S}=\mathscr{S}_{\phi,\gamma=0}^{\infty}-\mathscr{S}_{\phi,\gamma=1}^{\infty}$ is the interval in which the coherence-enhanced segregation applies. The coherence measurements and their associated scaling relations are discussed in the Supplemental Material. As shown in Fig.~\ref{fig:coherence}($b$), Eq.~\eqref{eq:Scoh} accurately captures the dependence of the steady-state degree of segregation on $\gamma$, including both the fully coherent and classical limits. Furthermore, when expressed in terms of $Pe_{\rm qca}$ and $\Pi_c$, results obtained from 84 simulations spanning different values of $\Gamma$, $\delta$, and $\gamma$ collapse onto the predicted master curve (Fig.~\ref{fig:coherence}$c$). The collapse demonstrates that coherence-enhanced segregation is governed by the interplay between segregation, mixing, and decoherence through two dimensionless parameters.

{\it Discussion.} We have presented an open quantum cellular automaton for particle-size segregation that combines coherent transport, dissipative segregation, and decoherence within a unified framework. Despite its minimal construction, the automaton reproduces the key features of classical segregation-diffusion dynamics without requiring ensemble averaging. Comparisons with experiments and continuum-theory solutions show excellent agreement in both steady-state concentration profiles and segregation degrees. Furthermore, the emergence of a quantum Péclet number that collapses the steady-state segregation degree demonstrates that the automaton recovers the classical segregation-diffusion balance.

More importantly, we identify a coherence-driven transport mechanism that enhances segregation beyond classical predictions. To isolate this effect, we initialize the automaton in a coherent superposition of normally graded and inversely graded states and systematically vary the dephasing rate $\gamma$. The resulting dynamics reveal that quantum coherence increases the degree of segregation while preserving the existence of segregation attractors, demonstrating that the classical segregation-diffusion balance does not necessarily define the maximum attainable segregation state. Introducing dephasing naturally leads to a second dimensionless coherent parameter which, together with the quantum Péclet number, governs the competition between segregation, mixing, and decoherence. Expressed in terms of these two parameters, the degrees of segregation obtained across a broad range of conditions collapse onto a single master curve. Within the automaton, coherent transport therefore acts as an additional transport channel that drives the system toward more strongly segregated steady states. More broadly, these findings suggest that transport mechanisms beyond classical diffusion may alter segregation dynamics and motivate the search for analogous coherence-like processes in particulate and many-body systems, as has been done in other classical systems \cite{Couder06,Plenio08,Bush15}. Whether similar transport enhancements can emerge in non-quantum granular materials through collective rearrangements, correlated motion, or other non-classical transport pathways is left open for future quantum-inspired experiments.

These findings establish quantum coherence as a mechanism capable of modifying segregation dynamics and demonstrate that open quantum systems can both reproduce and extend classical transport phenomena. More broadly, the proposed framework provides a route for exploring segregation, pattern formation, and transport processes in open many-body systems using quantum cellular automata.

The author would like to acknowledge the support from ANID through Fondecyt Iniciación 11240630 and the discussions with J. Acuña, J. Dumais and H. Ulloa. 
\bibliography{references}

\end{document}